\documentclass[a4paper,aps,pra,notitlepage,superscriptaddress,nofootinbib,reprint]{revtex4-1}

\usepackage{graphicx}
\usepackage{graphics}
\usepackage{color}
\usepackage{amsmath,amssymb,amsthm}

\newcommand{\be}{\begin{equation}}
\newcommand{\ee}[1]{\label{#1} \end{equation}}

\newcommand {\ket}[1]{\lvert \, #1\rangle}

\def\Tr{ {\textrm{Tr} }}

\def\f12{\frac{1}{2}}

\newcommand{\mean}[1]{\ensuremath{\left\langle #1 \right\rangle}}

\DeclareMathOperator{\arctanh}{arctanh}

\begin{document}

\title{Time Dilation in Quantum Systems and Decoherence: Questions and Answers}

\author{Igor Pikovski}
\affiliation{ITAMP, Harvard-Smithsonian Center for Astrophysics, Cambridge, MA 02138, USA}
\affiliation{Department of Physics, Harvard University, Cambridge, MA 02138, USA}
\author{Magdalena Zych}
\author{Fabio Costa}
\affiliation{Centre for Engineered Quantum Systems, School of Mathematics and Physics, The University of Queensland, St Lucia, QLD 4072, Australia }
\author{\v{C}aslav Brukner}
\affiliation{Vienna Center for Quantum Science and Technology (VCQ), University of Vienna, Faculty of Physics, Boltzmanngasse 5, A-1090 Vienna, Austria}
\affiliation{Institute for Quantum Optics and Quantum Information (IQOQI), Austrian Academy of Sciences, Boltzmanngasse 3, A-1090 Vienna, Austria}


\begin{abstract}
Recent work has shown that relativistic time dilation results in correlations between a particle's internal and external degrees of freedom, leading to decoherence of the latter.
In this note, we briefly summarize the results and address the comments and concerns that have been raised towards these findings.
In addition to brief replies to the comments, we provide a pedagogical discussion of some of the underlying principles of the work. This note serves to clarify some of the counterintuitive aspects arising when the two theories are jointly considered.
\end{abstract}

\maketitle
The interplay between quantum theory and general relativity offers many exciting and novel phenomena.
As shown in a series of recent works \cite{zych_quantum_2011, zych_general_2012, pikovskiuniversal2015}, time dilation causes entanglement between the center of mass of a quantum particle and its internal degrees of freedom. Conceptually, the effect can be understood as follows: a particle (e.g.\ an atom) is prepared in a state $\ket{\psi}$ of some internal degrees of freedom (e.g.\ optical energy levels).
The internal dynamics makes the internal state evolve to $\ket{\psi(\tau)}$, where $\tau$ is the proper time along the path of the particle. If the particle follows two different paths $\gamma_1$ and $\gamma_2$ in superposition, as for example in an interferometric experiment, the final state is $\ket{\gamma_1}\ket{\psi(\tau_1)} + \ket{\gamma_2}\ket{\psi(\tau_2)}$, where $\tau_1$, $\tau_2$ are the proper times accumulated along the two paths. One consequence is that this effect leads to decoherence of the particle's position, if the internal degrees of freedom constitute a sufficiently large ``bath'' \cite{pikovskiuniversal2015}. Related effects have been discussed in the context of matter-waves \cite{zych_quantum_2011, sinha2011atom}, photons \cite{zych_general_2012, brodutch2015post} and purely gravitational interactions \cite{gooding2015bootstrapping}, and experimentally simulated with a BEC \cite{margalit2015self}.

Despite the simplicity of the findings, and the fact that they follow from basic principles of quantum theory and relativity in a low-energy regime, our results have sparked several comments and, in a few cases, some skepticism. The objective of this note is to clarify the conceptual foundations of the results and to address some potential sources of confusion. This note is outlined as follows. In the first section we give a brief summary of the results of our work \cite{pikovskiuniversal2015}. In the second section, we briefly address comments on our work: the first comment \cite{ref:AB} deals with generic features of decoherence due to a finite bath, while the other two comments \cite{2015arXiv150705320B, 2015arXiv150705828D} purport to show a tension between our results and known physics (such as the equivalence principle), an incorrect claim rooted in a misunderstanding of the underlying physics.
In the third and last section we give a pedagogical overview of some of the concepts of quantum theory and relativity that enter our work, devised as a brief ``F.A.Q.'' that contains answers to frequently asked questions.
%
\section{Time Dilation and Quantum Superposition: Brief Summary}
Time dilation can be best understood by considering a system acting as an ``ideal clock''. In classical mechanics, this represents a point-like system, effectively following a single world line, with some internal dynamics that measures the passage of time. Let $L_{\text{rest}}$ be the Lagrangian describing the internal dynamics in the comoving frame. The action of the ``ideal clock'' is then given by
\begin{equation}
S=\int{L_{\text{rest}} d\tau},
\label{action}
\end{equation}
where $d\tau$ is the proper time element and the integral is taken over the classical world line of the particle. The expression for proper time for a post-Newtonian metric to lowest order in $1/c^2$ is given by
\begin{equation}
d\tau =\frac{1}{c} \sqrt{-g_{\mu \nu} dx^{\mu} dx^{\nu}} \approx dt \left(1+ \frac{\Phi(x)}{c^2} - \frac{v^2}{2c^2} \right) ,
\label{tau}
\end{equation}
where $t$, $x$ and $v=dx/dt$ are, respectively, the coordinate time, position and velocity of the system and $\Phi(x)$ is the gravitational potential. The time measured locally is always the proper time, but time dilation occurs between different world lines if their proper times are different. For instructive examples, it is sufficient to consider the external dynamics as fixed: the system follows a preassigned world line (or, in the quantum case which is discussed below, two or more preassigned world lines in superposition) and the internal dynamics evolves according to the world line's proper time. All classical tests of time dilation confirm this prediction of relativity.

In a general scenario, the evolution of the external degrees of freedom (i.e.\ the position of the particle) is not pre-assigned and has to be treated dynamically.
A useful approach is to derive the classical Hamiltonian corresponding to \eqref{action} by a Legendre transformation, although general covariance is less transparent in the Hamiltonian formulation. Using the metric \eqref{tau},
the Hamiltonian is
\begin{equation}
H = H_{\text{ext}} + H_0\left(1 + \frac{\Phi(x)}{c^2}  - \frac{p^2}{2m^2 c^2}\right),
\label{Hamiltonian}
\end{equation}
where $H_0$ and $H_{\text{ext}}$ are the internal and external Hamiltonians, respectively, and we have separated the constant mass contribution $mc^2$ from $H_0$ (see also \cite{pikovskiuniversal2015}, methods). For a free particle, $H_{\text{ext}}=mc^2 + p^2/2m + m \Phi(x)$ plus relativistic corrections. Any additional external forces acting on the system, such as those required to keep it at some height on earth or to perform an interferometric experiment, will contribute to $H_{\text{ext}}$ in \eqref{Hamiltonian}. The additional terms that couple $H_0$ to $x$ and $p$ are responsible for time dilation and are just a reformulation of eq.~\eqref{tau}.
The Hamiltonian \eqref{Hamiltonian} simply reproduces the effect of time dilation and captures its parametrization in given coordinates, in particular, its dependence on the position and velocity.
So for classical particles, Hamiltonian \eqref{Hamiltonian} captures some of the best-tested effects in modern physics. The coupling with momentum, $-H_0\frac{p^2}{2m^2 c^2}$, is simply the velocity-dependent special relativistic time dilation, while the coupling with position, $H_0\frac{\Phi(x)}{c^2}$, is the gravitational redshift.
Describing the dynamics on a fixed background space-time in terms of a Hamiltonian is standard procedure and higher order relativistic corrections can also be obtained within a consistent framework \cite{lammerzahl1995hamilton}.

The Hamiltonian treatment allows one to directly study how quantum systems behave in the presence of time dilation, by considering the canonically quantized dynamics. Instead of considering two test particles, we are interested in studying a single test particle that is in superposition on different world lines. Note that both, time dilation and the superposition principle have been extensively studied.
What has not been tested yet is the combination of the two; this provides one of the main motivations for investigating this regime. Using standard techniques to integrate out the internal degrees of freedom, one obtains the equation of motion for the particle's center of mass, described by a master equation, eq.~(5) in \cite{pikovskiuniversal2015}. This equation predicts the loss of coherence of particles in superposition along different world lines.

For pedagogical purposes, it is instructive to consider the superposition of only two, preassigned, world lines, which is also the paradigmatic example of a two-way interferometer. One can ask what happens to the coherence of such a superposition when the internal degrees of freedom are traced out. The net effect is that the interferometric visibility, a physical measure of coherence, is reduced according to eq.~(4) in \cite{pikovskiuniversal2015}:
\begin{equation}
V=\left|\Tr\left( \rho_0 e^{-i H_{0} \Delta \tau/\hbar} \right)\right|,
\label{visibility}
\end{equation}
where $\rho_0$ is the initial state of the internal degrees of freedom and $\Delta \tau$ is the proper time difference between the two world lines. This result  confirms the loss of coherence described by the Master equation.

For the simple example of a particle (with many internal degrees of freedom) being stationary but in a superposition of two different heights with height difference $\Delta x$, the loss of visibility is governed by a Gaussian decay with a characteristic decoherence time
\begin{equation}
\tau_{dec} = \frac{\hbar c^2}{\Delta H_0 g \Delta x} ,
\label{dectime}
\end{equation}
where $\Delta H_0 = \sqrt{\mean{H_0^2} - \mean{H_0}^2}$ is the internal energy variance. For a thermal state, this is directly proportional to the heat capacity. Using the Einstein model to get a simple dependence on the amount of internal degrees-of-freedom $N$ gives  $\Delta H_0 = \sqrt{3N}k_B T$.
%
\section{Published comments}
In this section we briefly address the comments on our work \cite{ref:AB, 2015arXiv150705320B, 2015arXiv150705828D}. The main concerns are the validity of the equivalence principle and whether the formalism we employ is justified. In short, the equivalence principle is fully respected and the effect we describe depends only on relativistic time dilation. The framework used is derived from relativistic quantum theory and describes the expected and known physics of low-energy systems in the presence of relativistic corrections. We stress that our results are based only on the validity of time dilation and low-energy quantum theory, with no assumptions about any new physical principles.
%
\subsection{Answer to Adler and Bassi. Or on the Quantum Poincar\'{e} Recurrence}
The note by Adler and Bassi \cite{ref:AB} agrees with our physical predictions, but contends that the decoherence induced by time dilation ``does not correspond to decoherence in the usual sense''  because the ``visibility does not approach zero for large times''. We note that this is in fact a generic and well understood feature of any decoherence model with a bath of finite volume, as also discussed in our manuscript. The fact that the visibility does not exactly vanish at infinite times is due to revivals of coherence, present in any physical decoherence model with a bath of finite volume \cite{ref:Bocchieri1956}. It is of little physical relevance for decoherence when the bath is sufficiently large, as the sheer amount of degrees of freedom to which the system is correlated renders such a revival to be practically unobservable. In the case of time dilation decoherence, the revival time already becomes much longer than the age of the universe for $\mu m$-superpositions and just a few dozen internal states with equally spaced frequencies.

The decoherence studied in our manuscript is of the same nature as any other decoherence source: It stems from correlations with other degrees of freedom which are integrated out. In the comment a comparison to collisional decoherence is made, but even there revivals of visibility will occur after sufficiently long times. This feature only disappears in the mathematical limit of strictly infinite volume as is implicitly assumed in \cite{ref:AB}. In fact, revivals have been confirmed experimentally in various setups and also in the context of light-matter interactions \cite{ref:Meunier2005}.

Various decoherence models all differ in terms of interactions and physical conditions, which results in different parameter and functional dependence of decoherence. The essential feature, common to all proposals, is that coherence is effectively lost for all practical purposes as a result of the process. Loss of coherence is induced by the inaccessibility of a vast range of degrees of freedom to which the system is correlated, independently of whether or not they become spatially separated from the system. For example, models of decoherence due to spin baths or quantum Brownian motion do not consider the environment to be far separated from the system.  In the limit of an infinite local environment, visibility vanishes for large times, just like in the infinite-volume limit for collisional decoherence. Therefore, time dilation or any other decoherence model with a local bath provides decoherence in the same way as collisional decoherence. What differs between different models of decoherence is the specific nature of the coupling. In our case it is proportional to internal energy, which causes the non-Markovian behavior and results in Gaussian decay in the specific limit discussed in our manuscript. A similar behavior is observed in other non-Markovian decoherence models, such as spin decoherence when coupled to the bath operator $\sigma_z$ \cite{ref:Cucchietti2005}. This is briefly discussed in sec.~\ref{Gaussian}.

The authors also re-derive our results for pure states. We highlight that the model studied in the comment is a special case of the formulas provided in our manuscript. The model considers pure states, as in our previous work \cite{zych_quantum_2011} which showed periodic loss and revivals of spatial coherence (see also sec.~\ref{twolevel}). The novel aspect of the model in \cite{ref:AB} is that it includes arbitrary number of energy eigenstates, as opposed to a 2-level system that we considered as an example in \cite{zych_quantum_2011}. Both are special cases of the generic treatment in our present manuscript \cite{pikovskiuniversal2015}, in particular they are contained in equation \eqref{visibility}. Also, the formulation of decoherence in terms of the energy variance is one of the main results reported in our work, resulting in the generic decoherence time \eqref{dectime}.
Importantly, our results are completely general: contrary to the claim in \cite{ref:AB}, neither the formula \eqref{visibility} nor the master equation assume anything about the composition of the system or the states and interactions of the internal degrees of freedom. This is a direct consequence of the universality of time dilation.
%
\subsection{Answer to Bonder, Okon, and Sudarski. Or on the Consistency of Time Dilation with the Equivalence Principle and Quantum Theory}
The comment by Bonder, Okon and Sudarski \cite{2015arXiv150705320B} claims that ``there must be something wrong with the results of [our] article'', arguing that our approach violates the equivalence principle and  contradicts Bargmann's superselection rule. The former claim is incorrect and  the latter does not apply to our work: the result of Bargmann holds for Galilei-invariant Hamiltonians, which is clearly not the case we consider. Below we clarify the basic aspects of classical relativity and quantum theory, related to the the mistakes in \cite{2015arXiv150705320B}.

\vspace{3pt}
i) The authors claim that
our results are wrong ``in the view of the equivalence principle'' and that in a uniformly accelerated frame ``there is no possibility for the claimed decoherence to occur, as there is simply nothing to attribute it to''.
The decoherence studied in our work is attributed to time dilation -- which does not contradict the equivalence principle. On the contrary, the equivalence principle sates that homogenous gravity and uniform acceleration are fully equivalent (as they are described locally by the same metric). They result in the same time dilation, and thus the same decoherence. For any experimental scenario, one can compute the time dilation along the superposed world lines to estimate the effect, as is clear from eq. \eqref{visibility}. The Hamiltonian derived is a consequence of the dependence of proper time on the particle's position and velocity. In our manuscript \cite{pikovskiuniversal2015} we discuss an example of a system in superposition between two fixed heights above earth, but the treatment is fully generic. The comment \cite{2015arXiv150705320B} invokes a different example where the superposed world lines are in free fall and ``an accelerated observer in a gravity-free region of space studies such a delocalized system''.
In sec.~\ref{covariance-dec} we solve an elementary exercise showing that even in this case, contrary to the authors' conclusions, an accelerated observer will measure time dilation and thus also decoherence.

\vspace{3pt}
ii) The authors further observe that systems in the presence of gravity will fall unless they are trapped and conclude that if the system is kept at a fixed height then ``it is clear that [the] result will depend on the exact nature of the interaction between the system and the external device and that it has nothing to do with gravity''.
In all experiments that measure gravitational time dilation the clocks have to be kept at a fixed position. Contrary to the authors' claim, time dilation is not caused by and does not depend on the exact nature of the trapping potential necessary to keep the clocks at fixed positions. Quite the opposite: different trapping potentials will result in the same measured time dilation if they keep the system at the same height, a key principle of relativity.
The force necessary to perform the experiment is explicitly taken into account in the master equation, eqs.\ (5) and (6) in the manuscript \cite{pikovskiuniversal2015}.
Note, however, that it is also possible to find mass configurations that will result in a gravitational potential difference between the locations of the amplitudes, but without any gravitational force \cite{hohensee2012force}. In such a case there will be time dilation -- and thus decoherence -- without any force (gravitational or non-gravitational) acting on the system. It has also been recently shown by Gooding and Unruh \cite{gooding2015bootstrapping} that time dilation can cause decoherence in the absence of any interaction other than gravity.

Regarding the role of gravity, decoherence due to gravitational time dilation can be considered a gravitational effect in precisely the same sense as gravitational time dilation from which it originates, see sec.~\ref{curvature}.

\vspace{3pt}
iii) The authors worry that ``the system's internal energy contributes to its effective mass [which] can have more than one value. (...) however, due to Bargmann's super-selection rule, ordinary nonrelativistic quantum mechanics cannot deal with such situations''.
The authors equate the use of a Hamiltonian to describe the dynamics with ``ordinary nonrelativistic quantum mechanics''. However, the Hamiltonian we use is derived from the framework of \textit{relativistic} quantum mechanics \cite{Dirac1958, berestetskii1974relativistic}  and contains relativistic corrections. It is the time-like component of the relativistic momentum four-vector and describes a particle on a Lorentzian space-time manifold. We are describing relativistic quantum mechanics in first quantization, where high-energy quantum field effects are negligible but where other relativistic effects still play a role. The Hamiltonian treatment is standard to describe relativistic corrections to single particles \cite{berestetskii1974relativistic, lammerzahl1995hamilton}.
Bargmann's result has  no relevance for our work, it says that superpositions of solutions to a non-relativistic Schr\"{o}dinger equation with different masses are not invariant under the Galilei group, and therefore unphysical in a Galilei-invariant theory in Euclidean space-time. No analogous result holds for relativistic quantum theory \cite{WeinbergQFT:1995}. Thus there is no mass-superselection rule when going beyond the Galilei group, as in our work\footnote{Coincidentally, the mass superselection is actually \textit{satisfied} in our specific example of a thermal internal state -- it is an incoherent mixture of states with different mass-energies, and thus also not excluded by the mass-superselection rule.} -- see also sec.~\ref{Bargmann}.
Note that the contributions of internal energy to the mass have also been previously considered for the case of a spin $1/2$ particle using the Dirac equation in curved space-time \cite{Morishima2004}.

Importantly, atoms can very well deal with superpositions of internal energies contributing to their mass -- as confirmed by 60 years of experiments in atomic physics. In particular, atomic clocks are precisely based on such superpositions and were recently utilized to measure time dilation in the group of D.\ Wineland \cite{chou2010optical}. If internal energy did not contribute to the mass for a superposition state, as the comment claims \cite{2015arXiv150705320B}, then atomic clocks would not be affected by time dilation, contrary to what is observed. Our treatment describes precisely situations such as in \cite{chou2010optical}, where quantum field theory effects can be neglected, but where relativistic corrections from the background space-time have to be included. (It is,  nevertheless, possible to formulate a field theory model giving the same results in the here considered regime, see methods in \cite{pikovskiuniversal2015}).

\vspace{3pt}
iv) The authors write that decoherence cannot explain the quantum-to-classical transition, since a subsystem is just a part of a larger, pure system, and thus constitutes an ``improper mixture''. While it is entirely correct that we consider subsystems of larger, pure systems, this is always the case within the field of decoherence and has no relevance for the inability to observe quantum coherence within the subsystem. When the subsystem is correlated with a large, effectively inaccessible environment, its coherence is for all practical purposes lost, see also sec.~\ref{classicality}.
For the study of how coherence  of one subsystem  is affected by correlations with another, only the reduced density matrix is relevant and not whether it stems from a proper or improper mixture. Our results do not hinge on an alleged physical difference between ``proper'' and ``improper mixtures'', which lies outside of quantum theory and the scope of our work.
%
\subsection{Answer to Di{\'o}si. Or the Coordinate Invariance of Time Dilation Decoherence}\label{diosi}

Di{\'o}si \cite{2015arXiv150705828D}, while agreeing that a spatial superposition will decohere in a gravitational field, finds ``paradoxical'' a reference-frame dependence of the effect and attempts to explain it as a signature of the ambiguity of the notion of center of mass. The apparent paradox, however, originate from a conceptual oversight. Rather than describing one and the same experiment from different reference frames, the author computes the effect for two physically distinct scenarios: equal-time measurements in an accelerated frame are compared with equal-time measurements in a free-falling one. These are different experiments and will not yield the same results, in accordance with the predictions of relativity and our work. Although not quite related with the rest of the argument, the ``paradoxical  frame-dependence'' is presented as an apparent tension with the equivalence principle. One should recall that the equivalence principle equates gravity with inertial effects in an accelerating frame; it does not imply equivalence of experiments in free-falling and accelerating frames, which would contradict basic physics. Being a direct consequence of relativity, decoherence due to time dilation is fully consistent with the equivalence principle, as also discussed in sec.~\ref{equivalence}.

When changing reference frames to describe one and the same experiment, two distant locations measured at an equal time in an accelerated reference frame appear as measured at different times in a free-falling frame, owing to the relativity of simultaneity. This is crucial when considering a particle in superposition of different locations from the point of view of different frames.  The coordinate-invariant expression \eqref{visibility} shows that the results of one and the same experiment are indeed reference-frame independent, as they should be. The description of the same experiment from the point of view of different observers is clarified in sec.~\ref{covariance-dec}.

In an attempt to ``explain the paradox'', it is pointed out in the comment \cite{2015arXiv150705828D} that accelerated and inertial observers assign  momenta to the particle that differ by $\left(m + H_0/c^2\right)v$ to leading order, where $v=gt$ is the relative velocity and $g$ is the acceleration. We remark that this is a direct consequence of the mass-energy equivalence: the canonical momentum includes not only the rest mass, but the total rest energy, including the internal energy $H_0$. In \cite{2015arXiv150705828D}, this is interpreted as an indication that ``the centre of mass canonical subsystem is ambiguous''.
While it is generally true that the center of mass is not uniquely defined in extended relativistic systems \cite{krajcik1974relativistic},
our entire analysis is in the limit where the extension of the system is negligible, as discussed in the methods section of \cite{pikovskiuniversal2015}. No issues related to the definition of the relativistic center of mass arise in this limit, as a single coordinate is assigned to the whole particle, and this is quite unrelated with the transformation properties of momentum.
What is relevant is the spatial superposition of the center of mass, but this does not affect the definition of the center of mass itself.

In conclusion, it is perhaps not unjustified to dub time-dilation decoherence ``paradoxical''. After all, the underlying physical effect -- time dilation -- is the same responsible for the famous ``twin paradox'' and shares similar counterintuitive features. With respect to reference frame dependence, we hope to have clarified that the observed effects do depend on what is measured and how it is measured, but not on the reference frame used to describe the measurement.
%
\section{F.A.Q.}
In this section we clarify some of the basic concepts behind our works \cite{zych_quantum_2011, zych_general_2012, pikovskiuniversal2015}. The selection of topics is based on the above comments and also on questions that have been raised in private communications and on conferences.
%
\subsection{Is the measurement of coherence reference frame dependent?}
No.
The coherence of a system is an observable quantity with a physical meaning within an experimental setup (such as a Mach-Zehnder interferometer).
For any specified experiment, the observed coherence will not depend on the reference frame, as guaranteed by the coordinate invariance of action \eqref{action} from which the effect is derived. For the case of two superposed world lines, decoherence only depends on the proper time difference between them, eq.~\eqref{visibility}, which is independent of the reference frame used to estimate it. The initial and final points of each world line have to be specified to make any meaningful statement about the total proper time. Decoherence does not depend on the reference frame but on the proper time difference between the amplitudes. Estimating the decoherence for various physical scenarios thus reduces to an exercise in classical relativity, computing the proper times along each interferometric path. The example considered in our work was a particle in superposition of being at two different, fixed heights \cite{pikovskiuniversal2015}. A different example is considered in the next section.
%
\subsection{Does a free-falling particle decohere?} \label{covariance-dec}
\begin{figure}[hb!]
	\centering
			\includegraphics[width=0.5\textwidth]{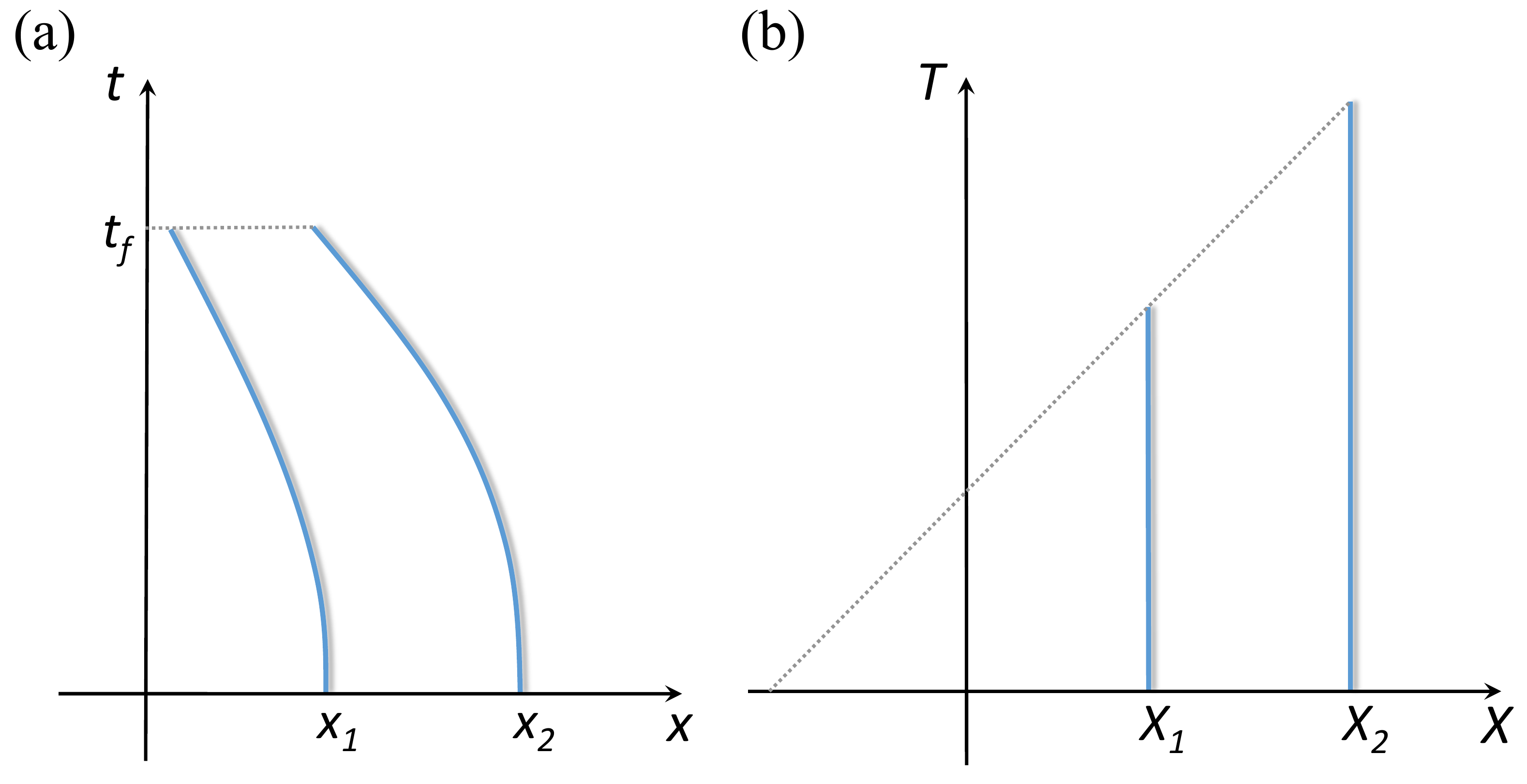}
		%
\caption{Two free-falling world lines as seen from the laboratory frame (a) and from a free-falling frame (b). The two world lines end on an equal-time plane in the laboratory frame (dashed lines), which appear as different times in the free-falling plane. The proper time difference between the world lines does not depend on the reference frame.}
\label{acceleratingpicture}
\end{figure}

Decoherence depends only on the overall time dilation between the superposed amplitudes. One can consider an experiment performed in a uniformly accelerated laboratory, but with the particle in free fall in superposition of different amplitudes (as in \cite{2015arXiv150705828D}). For simplicity, we consider here a laboratory accelerating in flat space-time, instead of one fixed above earth (the results coincide to leading order, in accordance with the equivalence principle). A particle is prepared at time $t=0$ in superposition of two different heights, $x_1$ and $x_2$. We assume that the spread of each wavepacket around each point is negligible and that no additional force is acting on the particle, so each wavepacket will follow a classical free-falling trajectory, see fig.~\ref{acceleratingpicture}a.
In order to estimate the coherence after some time $t_f$, we can rely on eq.~\eqref{visibility} and have to calculate the proper time along the two world lines. Notice that this is an exercise in classical relativity for which quantum mechanics plays no particular role: we can as well consider the proper time of two classical clocks following the two said world lines. It is convenient to introduce inertial coordinates:\
\begin{equation}
\begin{split}
t&= \frac{c}{g} \arctanh\left(\frac{c T}{X + c^2/g}\right), \\
x&= \sqrt{(X + c^2/g)^2- c^2 T^2} -c^2/g,
\end{split}
\label{rindler}
\end{equation}
where $X$, $T$ are Minkowski coordinates and $x$, $t$ Rindler coordinates \cite{rindler2012essential}. The two free-falling world lines are parameterized by Minkowski time $T$ and have constant Minkowski positions $X_1$, $X_2$, respectively (note that $x=X$ for $t=T=0$, so $X_j=x_j$, $j=1,2$), with proper time equal to Minkowski coordinate time. Importantly, since we are seeking the proper time difference for equal final \textit{Rindler time}, the two world lines have \textit{different} final Minkowski times (see fig.~\ref{acceleratingpicture}b),  yielding the proper time difference
\begin{equation}
\begin{split}
c \Delta \tau & = \Delta x \tanh (g t_f/c) \\
 & = \Delta x \left[g t_f/c +{\cal O}\left(g t_f/c\right)^3\right],
\end{split}\label{FallingInRindler}
\end{equation}
with $\Delta x = x_2-x_1$. Thus, for an accelerating observer the free-falling superposition appears to decohere. When looking at the same experiment from the point of view of a freely falling observer, the same proper time difference and thus the same loss of visibility is attributed. But in the free-falling reference frame the experiment considered does not test coherence \textit{at a given time}, rather the two amplitudes appear to be measured at different times.

One can also ask whether decoherence is observed for measurements that are synchronous in the free-falling frame. This is a different experiment and there is no reason to expect the same result as before. Clearly, for particles at rest in flat space-time and measured at equal times in Minkowski coordinates, no proper time difference and thus no decoherence is observed. However, one should not be led to conclude that no decoherence (or, equivalently, no time dilation) can be ever observed for systems in free fall (as should also be clear from the exercise above). In a recent work \cite{gooding2015bootstrapping}, Gooding and Unruh have shown that the effect discussed here can provide decoherence in systems that are only subject to gravity (and thus are, by definition, in free fall).

Note that the above example of a free-falling particle, which is the one referred to in the comments \cite{2015arXiv150705828D, 2015arXiv150705320B}, is not the example considered in our works \cite{zych_quantum_2011, zych_general_2012, pikovskiuniversal2015}, in which it is assumed that the particle is held in superposition at fixed heights. For a particle in superposition at fixed heights in Rindler coordinates, and measured at equal Rindler times, one finds the proper time difference $\Delta\tau=g\Delta x\, t^f/c^2$ confirming that decoherence will occur in this scenario. One can again ask whether decoherence is observed when the same accelerating particle is measured at equal times in a free-falling frame. We leave this as an exercise for the interested reader.

In the examples above the two world lines do not meet. It might perhaps be unclear how to assign a physical meaning to a measurement of coherence performed on distant wave packets. Typically, coherence is measured in interferometric experiments, in which the wave packets have to be overlapped on, say, a beam-splitter. This case is the one considered in our works \cite{zych_quantum_2011, zych_general_2012, pikovskiuniversal2015}, the superposed paths have common initial and final points (which may correspond to the first and second beam-splitter, or the source and detection points on a screen in a double-slit experiment).
No arbitrariness in the choice of measurement space-like planes arises in this case.
%
\subsection{Is the interaction Hamiltonian between internal and external degrees of freedom coordinate dependent?}\label{covariance}

Yes, this should not be surprising as a Hamiltonian is defined with respect to a given slicing of space-time in equal-time surfaces and so it is a coordinate-dependent object. The covariance of the effect is guaranteed by the fact that the Hamiltonian is obtained via Legendre transform from the action \eqref{action}, which is manifestly coordinate independent. For any given experiment, different observers will predict the same results, although they will in general use different Hamiltonians.

It can be useful to see explicitly how the interaction terms change when changing reference frame. A simple way to do this is to recall that the Hamiltonian is the $0$-th component of the $4$-momentum. The relation between $4$-momentum and rest energy is, in arbitrary coordinates,
\begin{equation}
g^{\mu \nu}p_{\mu}p_{\nu}= -(mc+ H_0/c)^2,
\label{4momentum}
\end{equation}
with the signature $(-+++)$ for the metric (see \cite{pikovskiuniversal2015}). The Hamiltonian in the given reference frame is then obtained by solving for $p_0$, which provides  $H \equiv cp_0$ as a function of $H_0$, $p_j$ and $x^{\mu}$ (through the position-dependent metric $g_{\mu \nu}$). The effective coupling between $H_0$ and $p_j$ and $x^{\mu}$  are then obtained by perturbative expansion\footnote{In the quantization of higher order terms, an ambiguity arises in the ordering of position and momentum operators. A definition of the Hamiltonian at arbitrary orders can be obtained by expanding the Klein-Gordon equation in curved space-time corresponding to eq.~\eqref{4momentum} to obtain a Schr{\"o}dinger-like equation with relativistic corrections \cite{lammerzahl1995hamilton}.}. Thus, the metric is all that is needed to know the form of the interaction $H_{int}$ in the given coordinates.

The coupling between position and internal energy in eq.~\eqref{Hamiltonian} arises from the expansion of $g_{00}(x)\sim -(1+2gx/c^2)$, where $g$ is earth's gravitational acceleration. In a free-falling frame, this coupling disappears, leaving only the coupling with momentum.
As a result, internal and external degrees of freedom will develop a different amount of entanglement in different reference frames, which brings us to the next question.
%
\subsection{Does the entanglement between internal and external degrees of freedom depend on the reference frame?}\label{covariance-ent}
Yes. As observed in sec.~\ref{covariance}, different observers use different interaction Hamiltonians and thus will observe different amounts of correlations (and in different bases) between internal and external degrees of freedom.
The key observation here is that a quantum state is defined at a given time on a given space-like surface. The state of a system ``at time $t$'' depends on whose time $t$ is considered. As soon as the described wave-packet has a non-negligible spatial extension, different observers will use different planes of simultaneity and thus will assign different states, simply because they are describing \emph{different physical situations}.
Given a particle in superposition at points $x_1$, $x_2$, a measurement of entanglement \textit{at a given time} for observer $A$
will, for observer $B$, appear as a measurement in which the two points are probed at different times, as discussed in detail in sec.~\ref{covariance-dec}. Thus the same measurement will not be interpreted by $B$ as a measure of the entanglement of the state.
Even though the definition of state, and thus the amount of entanglement, depends on the reference frame, different observers agree on the outcome of any measurement and in particular on the visibility of any interferometric experiment, which only depends on the proper time difference, eq.~\eqref{visibility}.

%
\subsection{Are the Predictions Consistent with the Equivalence Principle?}\label{equivalence}
Yes. Since the predictions are a direct consequence of relativistic time dilation, they automatically satisfy the equivalence principle, which requires that uniformly accelerated reference frames are  physically equivalent to those stationary in a homogeneous gravitational field. In particular, all results derived for accelerated observers are equivalent to those  derived for stationary observers on earth in the homogeneous field limit.

Moreover, one can easily see that the weak equivalence principle -- requiring that weight and inertia are equal -- is satisfied by the Hamiltonian \eqref{Hamiltonian}. The total rest energy  $mc^2+H_{0}$ gives both: inertia and weight. In particular, for the leading-order terms in the Hamiltonian \eqref{Hamiltonian}, we can see that the gravitational potential energy is $(m + H_0/c^2)\phi(x)$ (where $m\phi(x)$ is included in $H_{\text{ext}}$), which corresponds to a total weight $m + H_0/c^2$. Likewise, inertia is determined from the kinetic energy term: the non-relativistic term $p^2/2m$ together with the momentum coupling $-H_0\frac{p^2}{2m^2 c^2}$ is simply the first-order expansion of $\frac{p^2}{2(m + H_0/c^2)}$, which means that inertia is also equal to $m + H_0/c^2$.  Possible violations of the equivalence principle, which would indicate new physics, are discussed in ref. \cite{zych2015quantum}.
%
\subsection{As the effect does not depend on curvature, how can it be related to gravity?}\label{curvature}
As is clear from eq.~\eqref{visibility}, the decoherence effect depends on proper time and thus on the metric, not on curvature. A typical example of time dilation is one in which two clocks are held at fixed heights above earth. This effect is invariably referred to as ``\textit{gravitational} time dilation''. Similarly, the change in frequency as a light beam travels away from the earth's surface is commonly known as ``\textit{gravitational} redshift''. Such effects depend on the $g_{00}$ component of the metric and not on curvature. In a first-order approximation, $g_{00}\sim -\left(1+2\phi(x)/c^2\right)$, where $\phi(x)$ is identified with Newton's potential. In the same way, the decoherence of a particle held in superposition at two heights above earth depends on $g_{00}$ and is as much related to gravity as gravitational time dilation or redshift. Note that gravity is not necessary for time dilation: a difference in velocities between superposed paths will also lead to time dilation and thus decoherence. Decoherence due to gravitational time dilation is simply a special case of decoherence due to time dilation.
It is of course legitimate to consider only curvature-related effects as gravitational.
Accordingly, gravitational redshift, gravitational time dilation, falling apples, and also decoherence due to time dilation should not be understood as related to gravity. It should be clear that this is rather a semantic issue that does not affect the physical predictions.
%
\subsection{Is it consistent to treat relativistic quantum effects without using quantum field theory?}
Any physical model, at least any known to date, has a limited range of applicability. For example, quantum field theory in curved space-time cannot describe the back-action of mass-energy on space-time at arbitrary energy scales. In our work, we considered low-energy quantum systems such as atoms, molecules, nanospheres, etc., in a regime in which high-energy quantum field effects, such as particle creation/annihilation, can be neglected. This regime can be well approximated by relativistic quantum mechanics in first quantization. Although it has a limited range of applicability, relativistic quantum mechanics is a well-understood framework that yields powerful predictions, such as corrections to atomic spectra \cite{berestetskii1974relativistic, BjorkenDrell:1964B}. General relativistic effects can also be included in a first-quantization treatment, for example by including corrections to the non-relativistic Schr{\"o}dinger equation \cite{lammerzahl1995hamilton}. The novelty of our analysis is to consider the effects arising from the internal structure of the quantum particle, which can be incorporated in the Klein-Gordon equation by adding the internal energy contribution to the mass.

Of course, it should also be possible to derive the same effects within the framework of quantum field theory in curved space time. An example of such a derivation, based on a simple model, was presented in the methods section of our work \cite{pikovskiuniversal2015}.
%
\vspace{-4pt}
\subsection{How can the mass-energy equivalence and the superposition principle be reconciled with the mass superselection rule?}
\label{Bargmann}
They do not have to be reconciled -- they belong to different physical theories (similarly, one does not need to reconcile the notion of absolute time in Galilean mechanics with relativistic time dilation). The mass superselection rule is a result in \textit{non-relativistic} quantum mechanics. It originates from the non-commutativity of the  generators of the boost and translation in the Lie algebra of the representation of the Galilei group on the space of solutions to the non-relativistic Schr\"{o}dinger equation. Specifically, in one space dimension, the Galilei boost generator is $\hat K= m\hat X-\hat Pt$, where $\hat X, \hat P$ are the position and momentum, respectively, and $m$ is the mass of the particle; the generator of translations is $ \hat P$. Thus, $[\hat K, \hat P]=i\hbar m$, whereas in the Lie algebra of the  Galilei group itself these generators commute. Such a representation  is called projective and results here in an additional phase factor proportional to the mass, which in turn leads to the mass-superselection rule: denote by $g$ and $g'$ the Galilei group elements of a spatial translation by $a$ and a boost by $v$, respectively. They satisfy $g^{\prime -1} g^{-1} g' g = 1$ (identity element of the Galilei group). However, their representations as operators on the Hilbert space, $\hat U_g=e^{-i\hat Pa/\hbar}$ and $\hat U_{g'}=e^{-i \hat{K}v/\hbar}$, satisfy $\hat U_{g'}^{-1}\hat U_{g}^{-1}\hat U_{g'}\hat U_{g}=e^{-imva/\hbar}\hat I$.
Applying this sequence to a superposition of states characterized by different masses $m$ and $m'$ results in a relative phase $e^{iva(m-m')/\hbar}$ and therefore a different physical state, unless $m = m'$. However, this operation should represent identity of the Galilei group and cannot alter physical states. Hence a superposition of states with $m\neq m'$ is considered unphysical in a Galilei invariant theory and is thus ``forbidden'' -- this is the original argument of Bargmann behind  the superselection rule for the mass.  In contrast,
representations of the Poincar\'{e} group in relativistic quantum theory have the same Lie algebra as the group itself and thus they are unitary -- i.e.\ not projective -- representations of the Poincar\'{e} group. In particular, there is no superselection rule for the mass in relativistic quantum mechanics \cite{WeinbergQFT:1995}. So if one does not insist on Galilei invariance, superpositions of states with different mass-energies are not ``forbidden''.
(This can also be understood by simply recalling that relativistic quantum mechanics has to incorporate both: mass-energy equivalence of relativity and the superposition principle of quantum theory. As a result superpositions of internal energies must contribute to the mass in precisely the same way as the eigenstates alone.)
%
\subsection{What happens to particles with few internal degrees of freedom?} \label{twolevel}
In our previous works \cite{zych_quantum_2011, zych_general_2012}, we studied pure internal states which evolve between mutually distinguishable states. In particular, as a specific example in \cite{zych_quantum_2011} we considered a two-level internal state $\ket{\psi}=\frac{1}{\sqrt{2}}\left( \ket{e} + \ket{g}\right)$,  which causes periodic loss and revivals of coherence of the spatial degrees of freedom:
\begin{equation}
V = \left| \textrm{ cos}\left( \frac{\omega \Delta \tau}{2}\right)\right| ,
\label{visibility2lvl}
\end{equation}
where $\omega$ is the transition frequency between the internal ground $\ket{g}$ and excited $\ket{e}$ states. The above result can also be derived directly from eq.~\eqref{visibility}. For the example of fixed vertical separation $\Delta x$ in the homogeneous gravitational field $g$ kept for (laboratory) time $t$, the proper time difference is $\Delta \tau = g \Delta x t/c^2$.

In a recent experiment, the above predictions have been simulated in a BEC interference setup in the presence of inhomogeneous magnetic fields \cite{margalit2015self}.
%
\subsection{In what sense is the effect universal?}
The coupling of position and momentum with the internal energy $H_0$, eq.~\eqref{Hamiltonian}, does not depend on the nature of the binding energies and interactions that define $H_0$, which can describe any internal dynamics. This is because the coupling is a consequence of time dilation, which does not depend on the construction of the clock used to measure time. Therefore, decoherence due to time dilation is as universal as time dilation itself and affects any composite quantum system.
%
\subsection{Is coherence not going to be restored after the beams are brought back together?}
Coherence depends on the total elapsed proper time of each of the interfering amplitudes. Coherence can be predicted once an experimental setup is fully specified, i.e.\ the time between preparation and measurement of a superposition, as well as the heights and velocities of each interfering amplitude, such that the total proper times can be computed (see also sec.~\ref{covariance-dec}). The coherence will of course change if the experiment is changed. For example, if a setup is prepared in such a way that in the end no time dilation occurs, then no decoherence due to time dilation will take place (for example, by canceling the difference in proper time from the gravitational field by appropriately changing the velocities). All this is described by eq.~\eqref{visibility}, and also equivalently by the full master equation in \cite{pikovskiuniversal2015}.
Needless to say, the reduced state of the system still takes into account how the bath dynamically affects the coherence. An example of this is eq.~\eqref{visibility2lvl}, where the coherence changes periodically as the internal two-level system periodically evolves from identical to orthogonal states along the two superposed paths.

Another possible source of confusion, which might lead to the intuition that coherence is always going to be restored when the superposed paths are brought together, comes from the example of a charged particle in an interferometer. In the middle of the interferometer, the particle is in superposition of different positions and, for each position, it produces a different electric field. Thus, ``on the fly'', the position of the particle is entangled with the field and so appears ``decohered'' when the field is traced out. However, when the different amplitudes interfere, the particle is again in a single place and the entanglement with the field disappears \cite{unruhfalse2000}.

For a particle generating a Newtonian potential, the situation would be the same. However, the effect we describe is not related to the gravitational field of the particle itself: decoherence depends on the proper time difference accumulated by the different amplitudes in superposition, see eq.~\eqref{visibility}. Such a proper time difference does not have to vanish for two paths with the same starting and end points, as experiments confirm \cite{hafele1972observed}. For a particle following trajectories with a sufficient proper time difference coherence is not restored and a net visibility loss will be observed.
%
\subsection{Why does the decoherence follow a Gaussian decay and not the usual exponential decay?} \label{Gaussian}
This is a direct consequence of the specific interaction Hamiltonian, eq.~\eqref{Hamiltonian}. In all models of decoherence, the system degrees of freedom couple to some degrees of freedom of the bath with an interaction Hamiltonian $H_{int}$. The equation of motion for the system of interest, in the interaction picture and Born approximation, is
\begin{equation}\label{ME}
\dot{\rho}_s(t) = - \frac{1}{\hbar^2} \int_0^t dt' \Tr_B\left[H_{int}(t), \left[H_{int}(t'), \rho_s(t') \otimes \rho_B \right] \right]
\end{equation}
where the trace is taken over all bath degrees of freedom. Writing the interaction Hamiltonian as $H_{int} \propto S \otimes B$ for some system operator $S$ and bath operator $B$, the relevant quantities in the above equation are the bath auto-correlation functions in the interaction picture, $\mean{B(t)B(t')}= \Tr_B \left[ B(t)B(t') \rho_B \right]$. In many models of decoherence, these decay very rapidly such that one can approximate $\mean{B(t)B(t')} \propto \delta(t-t')$. For example, in quantum Brownian motion \cite{caldeirapath1983} one has $B = \sum_i a_i X_i$, i.e.\ coupling to the positions of the bath degrees of freedom with coupling strengths $a_i$, which yields in the high temperature limit $\mean{B(t)B(t')} \approx 4 m \gamma k_B T\delta(t-t')$, where $m$ is the mass, $T$ the temperature and $\gamma$ the damping coefficient of the system.

In contrast, time dilation causes a coupling to the internal energy of the system, $B = H_0$. Thus the bath auto-correlation functions in the interaction picture remain constant $\mean{B(t)B(t')}= \mean{\left(H_0 - \bar{H}_0\right) ^2}$, where $\bar{H}_0 = \mean{H_0}$ is the mean internal energy (see also methods in \cite{pikovskiuniversal2015}). This decoherence is therefore in the exact opposite limit than the Markovian models. A few other models are of this form, as for example the one by Cucchietti, Paz and Zurek \cite{ref:Cucchietti2005} in which a spin couples to the bath spin operators $B=\sum_i \sigma_z^{(i)}$, and which also results in a Gaussian decay of coherence.
%
\subsection{Does decoherence explain classicality?} \label{classicality}
The so-called transition to classicality is a debated issue in the philosophy and foundations of quantum mechanics \cite{ZurekDecoherence2003, schlosshauer2007decoherence}. One of the sources of controversy is the very definition of classicality and in which sense the transition is to be understood. One can restrict the discussion to practical questions of why effects such as quantum interference \cite{eibenberger2013matter} or violations of Bell inequalities \cite{freedman72} are not observed on everyday scales. Decoherence explains this fact by considering interactions with an environment which causes a suppression of coherence in a so-called ``pointer basis'', such that typical quantum effects cannot be observed.
This follows directly from quantum theory and the study of open quantum systems.

A further problem often associated with the quantum-to-classical transition is the so-called measurement problem: the questions why a specific outcome of a quantum measurement occurs and what constitutes a measurement. As an inherently probabilistic theory, quantum mechanics does not provide any means to explain the occurrence of specific outcomes beyond predicting the probability with which they occur. In fact, any theory providing better predictability than quantum mechanics would have to be non-local \cite{colbeck_no_2011} and contextual \cite{PhysRevA.83.042110}. Thus, quantum theory or decoherence cannot answer this aspect of the measurement problem. The study of what exactly constitutes a measurement is not addressed by decoherence and is beyond the scope of our work.
%
\subsection{Is the effect described equivalent or related to Penrose's gravitationally-induced collapse of the wave function?}
No. The suggestion of Penrose \cite{penrose1996gravity} is based on an inherent modification of quantum theory. It is argued that the superposition principle breaks down if the systems reach a sufficient size. The philosophical motivation for this suggestion is the consideration of superpositions of different metrics. However, the gravitationally-induced collapse of the wave function does not follow from quantum theory and is a speculative modification thereof. In contrast, our result stems fully from within quantum theory. The dynamics we consider is unitary for the total system -- the decoherence takes place due to correlations with an environment (the internal degrees of freedom of the particle), which are induced by time dilation.
%
%
\section{Acknowledgments}
We thank Markus Arndt, Markus Aspelmeyer, Angelo Bassi, Lajos Di\'{o}si, Fay Dowker, Cisco Gooding, Daniel Greenberger, Eduardo Martin-Martinez, Holger M\"{u}ller and Carlo Rovelli for discussions. This work was supported by the NSF through a grant to ITAMP, the Austrian Science Fund (FWF) through the Special Research Program Foundations and Applications of Quantum Science (FoQuS) and Individual Project (No.\ 2462), the FQXi, the Australian Research Council Centre of Excellence for Engineered Quantum Systems through grant number CE110001013 and by the Templeton World Charity Foundation, grant TWCF 0064/AB38.

\bibliographystyle{linksen}
\bibliography{Pikovski_QA}


\end{document}